\documentclass[opinion]{sigirforum}

\def\pubissue{Pre-Print}

\usepackage{fancyvrb}
\usepackage{fvextra}
\usepackage{caption}

\begin{document}
\title{From AutoRecSys to AutoRecLab: A Call to Build, Evaluate, and Govern Autonomous Recommender-Systems Research Labs}

\authors{
\author[joeran.beel@uni-siegen.de]{Joeran Beel}{University of Siegen}{Germany}
\and
\author[bela.gipp@uni-goettingen.de]{Bela Gipp}{University of Göttingen}{Germany}
\and
\author[tobias.vente@uantwerpen.be]{Tobias Vente}{University of Antwerp}{Belgium}
\and
\author[moritz.baumgart@student.uni-siegen.de]{Moritz Baumgart}{University of Siegen}{Germany}
\and
\author[philipp.meister@student.uni-siegen.de]{Philipp Meister}{University of Göttingen}{Germany}
}

\maketitle
\begin{abstract}
Recommender-systems research has accelerated model and evaluation advances, yet largely neglects automating the research process itself. We argue for a shift from narrow AutoRecSys tools—focused on algorithm selection and hyper-parameter tuning—to an Autonomous Recommender-Systems Research Lab (AutoRecLab) that integrates end-to-end automation: problem ideation, literature analysis, experimental design and execution, result interpretation, manuscript drafting, and provenance logging. Drawing on recent progress in automated science (e.g., multi-agent "AI Scientist" and "AI Co-Scientist" systems), we outline an agenda for the RecSys community: (1) build open AutoRecLab prototypes that combine LLM-driven ideation and reporting with automated experimentation; (2) establish benchmarks and competitions that evaluate agents on producing reproducible RecSys findings with minimal human input; (3) create review venues for transparently AI-generated submissions; (4) define standards for attribution and reproducibility via detailed research logs and metadata; and (5) foster interdisciplinary dialogue on ethics, governance, privacy, and fairness in autonomous research. Advancing this agenda can increase research throughput, surface non-obvious insights, and position RecSys to contribute to emerging Artificial Research Intelligence. We conclude with a call to organise a community retreat to coordinate next steps and co-author guidance for the responsible integration of automated research systems. 

\end{abstract}

\section{Introduction}
The Recommender Systems (RecSys) community is approaching its 20th anniversary. In the past two decades, the community  has consistently demonstrated its ability to adapt to new technologies and challenges. Yet, while our community continually advances algorithms and evaluation techniques and explores novel areas such as LLMs for generating recommendations, we have largely overlooked a transformative paradigm shift: automating the research process itself. In broader machine learning (ML), automated machine learning (AutoML), and emerging Artificial Research Intelligence (ARI) systems are beginning to conduct research autonomously – generating hypotheses, running experiments, and drafting papers with minimal human input \citep{ lu2024ai}. The RecSys field, by contrast, is lagging behind. Only a few rudimentary Automated Recommender Systems (AutoRecSys) tools exist (e.g. Auto-Surprise\citep{anand2020auto}, LensKit-Auto\citep{vente2023introducing}, LibRec-Auto\citep{sonboli2021librec}), mostly limited to algorithm selection and hyper-parameter tuning, often as one-off prototypes. We believe the time has come for the RecSys community – including researchers, conference organizers, industry practitioners, and tool developers – to fundamentally rethink how we conduct our research. We must embrace the vision of a fully Autonomous Recommender Systems Research Lab (AutoRecLab) that goes far beyond existing AutoRecSys, and we must act now to avoid being left behind.

\section{Related Work (AI Research Agents)}
In recent years, bold strides in automating scientific research have shown what’s possible and what is yet to come. A salient example is Sakana’s AI Scientist, a system that “automates the entire research lifecycle” – from idea generation and experiment design to result analysis, paper writing, and even peer review\citep{lu2024ai}. Sakana’s AI Scientist v1 (2024) and the improved v2 (2025) illustrate the rapid progress in this arena. While our recent evaluation of the AI Scientist revealed notable shortcomings, the evaluation also proved to be an eye-opener \citep{beel2025evaluating}. Despite its limitations, the AI Scientist achieved a significant leap forward in research automation, producing complete research manuscripts with minimal human intervention and challenging conventional expectations of AI-generated science \citep{beel2025evaluating}. In fact, it is nowadays hard -- or near impossible -- to distinguish an AI-generated manuscript from work generated by "an unmotivated undergraduate student rushing to meet a deadline" \citep{beel2025evaluating}. Equally astonishing is the efficiency – the AI Scientist can generate a full research paper for less than \$20 in compute costs and a few hours of human labour at most \citep{lu2024ai,beel2025evaluating}. Just a few years ago, fully autonomous research was virtually nonexistent; today it is a reality in prototype form, signalling that achieving Artificial Research Intelligence (ARI) – AI agents that conduct research at a near-human level – is within sight.

Other AI-driven research systems reinforce this trend. Google recently introduced an “AI Co-Scientist,” a Gemini 2.0–based multi-agent system that proposes novel scientific hypotheses and research plans as a virtual collaborator\citep{gottweis2025towards}. Google’s AI Co-Scientist is explicitly designed to accelerate discovery by automating hypothesis generation and experimental design – essentially taking on parts of a scientist’s role. Early results in biomedical domains show the AI Co-Scientist formulating promising drug candidates and research insights\citep{gottweis2025towards}. The very notion that a major research lab now touts an AI “co-scientist” underscores how serious and mainstream automated research is becoming.

Meanwhile, Sakana’s AI Scientist v2 demonstrated its prowess by producing the first fully AI-generated research paper to pass peer review at a top-tier ML workshop \citep{yamada2025ai}. In an experiment at ICLR 2025, the AI Scientist-v2 autonomously wrote and executed three papers; one was accepted by reviewers -- who were unaware it was machine-generated -- as a valid scientific contribution\citep{sakana2025aiscientistv2}. Although the paper was ultimately withdrawn to avoid ethical concerns, its acceptance marked a historic milestone: for the first time, an AI agent’s research was judged worthy of publication by peer scientists\citep{sakana2025aiscientistv2}. These developments, while outside the traditional RecSys sphere, carry an urgent message: the automation of research is accelerating, and we as a community must catch up.

Recent efforts illustrate that SakanaAI’s AI Scientist and Google’s AI Co-Scientist are just two among many emerging systems targeting automated scientific research. Several frameworks now orchestrate multiple stages of the research lifecycle with minimal human input. For example, AI-Researcher is a fully autonomous research agent that spans literature review, hypothesis generation, experiment implementation, and manuscript drafting \citep{Tang2025}. Similarly, Agent Laboratory progresses from a human-given idea through automated literature analysis, experiment execution, and report writing, yielding both code and a scientific paper as outputs \citep{Schmidgall2025}. Other systems emphasize end-to-end data-driven discovery: Data-to-Paper guides collaborating LLM agents from raw dataset exploration to a complete, human-verifiable research article \citep{Ifargan2024}. In the same vein, CodeScientist integrates idea generation with code-based experimentation, conducting hundreds of autonomous experiments and assembling results into publishable findings \citep{Jansen2025}. Some approaches explicitly encode scientific methodology – AIGS (AI-Generated Science), for instance, introduces a falsification agent to rigorously test and refine hypotheses in an entirely agent-led research loop \citep{Liu2024}.

Efforts have also specialized or extended the paradigm. In the biomedical domain, Virtual Lab demonstrated that multi-agent systems can design novel experimental candidates: Swanson et al. showed LLM-based “scientists” collaborating (with minimal human guidance) to propose and experimentally validate new SARS-CoV-2 nanobodies \citep{Swanson2024}. Other work has explored enabling collaboration among AI scientists themselves. AgentRxiv provides a shared preprint server where autonomous research agents publish and retrieve interim results, allowing multiple agent laboratories to build on each other’s findings over iterative cycles \citep{SchmidgallMoor2025}. 

Together, these systems underscore a growing trend toward automating large portions of the scientific process. They highlight that Google’s and Sakana’s systems are part of a broader movement: a range of architectures now attempt to generate ideas, run experiments, interpret outcomes, and even draft papers autonomously, marking significant steps toward AI-augmented or even AI-driven science.

\section{From AutoRecSys to AutoRecLab}
Current efforts in RecSys automation are clearly too narrow in scope. AutoRecSys tools like Auto-Surprise, LensKit-Auto, and LibRec-Auto – to name a few – focus primarily on optimizing algorithm choice and parameters within existing pipelines. They ease repetitive tasks for an “inexperienced user” building a recommender system. And indeed, AutoRecSys tools may outperform default human-chosen baselines \citep{vente2025potential}. Yet, these tools barely scratch the surface of what could be automated. They do not generate new research questions, design novel experiments, or write up findings. In short, they optimize within a human-defined sandbox, whereas the emerging ARI systems aim to expand the sandbox itself. 

Moreover, existing AutoRecSys libraries are underdeveloped – often academic prototypes without the robustness, integration, and community support that mature AutoML frameworks enjoy\citep{vente2025potential}. The RecSys community’s prevailing mindset has been that research automation, or more precisely, the support of research and development, is limited to automated hyper-parameter tuning or trivial experiment scripting. This view is short-sighted. As the AI Scientist and other tools illustrate, virtually every step of the scientific process – including idea generation, literature review, experimental design, result interpretation, and paper writing – can be at least partially automated\citep{yamada2025ai}. So far, the RecSys community has been optimizing individual models and recommender algorithms, but not the process of researching and inventing those models.

The hesitation is understandable: recommender systems research often involves nuanced human factors, domain-specific knowledge (e.g. user behavior theories), and complex experimental protocols (online A/B tests, privacy considerations, etc.). Some may argue that these aspects make RecSys research uniquely hard to automate. However, similar arguments could be (and have been) made about other fields. If an AI can propose a new biomedical experiment or suggest a novel deep learning architecture, there is little fundamental reason it could not propose a new recommendation algorithm or experimental methodology. Indeed, elements like hyper-parameter tuning or simulation of user interactions are more straightforward to automate than creative insight – yet we see AI beginning to demonstrate glimmers of creativity in research. The real barrier in RecSys is not the impossibility of automation, but the lack of organized effort and vision in pursuing it.

We believe the RecSys community needs to expand its ambition. We should envision an “AutoRecLab”: an integrated system or collection of interoperable tools that can autonomously perform end-to-end research in recommender systems. Imagine an AI agent that can ingest the entire RecSys literature, identify open problems or generate new ideas (e.g. a fresh approach to fairness in recommendations or a hybrid of content-based and collaborative filtering never tried before), design appropriate experiments on standard benchmarks or simulations, iteratively optimize algorithms, and then produce a draft paper complete with evaluations and references. This AutoRecLab would not necessarily operate in isolation – it could work with human researchers in a loop, speeding up tedious tasks and perhaps offering non-obvious insights. Crucially, it would also maintain rigorous records of its process (akin to a research log or “lab notebook” for AI) so that humans can audit and learn from the automated research process \citep {beel2025evaluating}. Realizing this vision is a grand challenge, but one that RecSys as a field is well-positioned to tackle, given our expertise at the intersection of algorithms, data, and user-centric evaluation.

We, the RecSys community, should care about autonomous research not only because it can make our work more efficient, but because an AutoRecLab is, at its core, a recommender system; our recommender-systems methods and theory are directly useful to build such a lab, and other disciplines will benefit from RecSys technologies when developing autonomous science agents. Autonomous science can be framed as a sequential recommendation problem: at every step, an agent must rank and select “items” (hypotheses, datasets, methods, experimental designs, analyses, writing moves) conditioned on context (prior literature, results so far, constraints such as cost, time, carbon, and ethics) and feedback (measured effect sizes, robustness checks, peer review). The core phenomena are recommender-systems phenomena: exploration–exploitation under uncertainty; multi-objective ranking (novelty, validity, reproducibility, impact); cold-start (new domains and tasks); counterfactual and off-policy evaluation (estimating payoffs of unrun experiments); provenance and explainability (research logs as recommendation rationales); and online evaluation (benchmarks, competitions) versus offline metrics. 

Under this lens, an autonomous researcher or an autonomous researcher lab is a large, modular recommender system whose sub-recommenders propose problems, select baselines and hyperparameters, schedule experiments, prioritize analyses, and iteratively revise manuscripts, all governed by logged policies for audit and attribution. This perspective both motivates RecSys-native methods (bandits, slates, constrained optimization, causal inference) for automated research and clarifies why building, evaluating, and governing AutoRecLab belongs squarely within the RecSys agenda.

\section{A Call to Action for the RecSys Community}
Moving toward an AutoRecLab paradigm will require concerted effort across our community. We call on RecSys researchers, industry labs, conference organizers, and platform builders to jointly advance the following agenda:

\subsection{Invest in AutoRecLab Prototypes}
We should initiate and fund projects that develop prototype systems integrating multiple stages of the research process. This goes beyond AutoML-for-RecSys. For example, an academic–industry collaboration could create an open-source “RecSys Research Assistant” that generates problem hypotheses (e.g. new recommendation scenarios or loss functions), suggests experiment designs (which datasets, which baselines, etc.), runs those experiments, and assembles results into report drafts. Even if early versions are clumsy, they will be invaluable for identifying challenges and spurring improvement. Similar prototypes in other fields (e.g. Sakana’s AI Scientist) have shown surprising capabilities at low cost, so RecSys should start building its own. We particularly encourage leveraging large language models (LLMs) fine-tuned on RecSys literature to power the idea generation and writing components, and integrating AutoML techniques for experiment execution. In the first step, such prototypes might be an ensemble of "\textit{Automated} Recommender-Systems Research Labs": humans still oversee the majority of the work, and the tools simply automate large parts of the research pipeline. In the long run, we envision "\textit{Autonomous} Recommender-Systems Research Labs" in which AI agents work mostly independently.

\subsection{Establish Benchmarks and Competitions}
To catalyze progress, the community should establish benchmark tasks and datasets specifically for automated RecSys research agents. For example, a benchmark could require an AI agent to take a standard RecSys dataset (MovieLens, Amazon reviews, etc.), develop a novel recommendation approach, and produce a results report with minimal human help. We propose organising competitions or “AI Researcher” challenges at RecSys and related conferences, where different automated agents compete to produce the best recommendation models or even the best research paper on a given theme. This competitive setting, akin to the general RecSys challenges, would greatly accelerate development and highlight the benefits of RecSys automation. In fact, the idea of AI-generated research competitions has already been floated in the IR community \citep{beel2025evaluating} – we suggest RecSys take the lead in making it a reality.

\subsection{Embrace AI-Generated Submissions (Thoughtfully)}
Our conference organizers and reviewers should not dismiss AI-generated research out of hand, but rather create space to evaluate it rigorously. We encourage setting up sandbox tracks or workshops where fully or partially AI-generated papers can be submitted (with transparent disclosure) and reviewed. This “Turing Test” for RecSys research would be enlightening: if an AI can produce a paper that seasoned reviewers find acceptable, that is a result we must pay attention to (as happened in the ICLR 2025 experiment\citep{sakana2025aiscientistv2}). Even if such papers are flawed, reviewing them will help develop criteria for evaluating AI-generated science and spur improvements in both AI systems and our review process. Eventually, we may see a day when an AI-written paper on recommender systems is accepted in the main conference – we should prepare for that possibility rather than ignore it.

\subsection{Develop Standards for Attribution and Reproducibility}
As we incorporate AI agents into the research pipeline, we need new standards for transparency and credit. We echo prior suggestions for standardized research logs and metadata that record each step an AI research agent takes\citep{beel2025evaluating}. Such logs (akin to detailed experiment provenance records) would ensure reproducibility and allow humans to trace how an AI arrived at a result or a conclusion. They also help assign attribution – for instance, distinguishing which parts of a paper were written or decided by an AI vs. a human. The community could develop a simple markup language or schema for documenting AI contributions in research artifacts \citep{beel2025evaluating}. By establishing these norms early, we can avoid future ethical ambiguities and ensure that AI-augmented research remains transparent and trustworthy.

\subsection{Foster an Interdisciplinary Discussion} 
We urge RecSys researchers to actively engage with the broader movement of automating science and affected disciplines. There should be cross-pollination between RecSys and teams working on automated science in other fields. Workshops, panels, and interdisciplinary collaborations (e.g. with the IR, ML, or science-automation communities) will be crucial. This is not only to share technical approaches, but also to shape the policy and ethical guidelines around AI-conducted research. Questions about the role of human creativity, accountability for AI-generated results, and the impact on researcher training are looming large \citep{heath2025aiready}. The RecSys community should be at the table in these discussions, helping to ensure that as we automate parts of research, we do so in a way that benefits science and society. By proactively engaging, we can also address domain-specific concerns – for example, how would an AutoRecLab account for user privacy and fairness when autonomously experimenting with recommender algorithms?

\subsection{Participatory Strategic Retreat}
The concepts presented in this paper are intended as an initial impulse for reflection and collective dialogue. To advance the vision of fully autonomous research labs, the involvement of key actors across the RecSys ecosystem is essential — from established leaders in academia and industry to emerging researchers whose careers will unfold alongside increasingly autonomous research tools. We propose that a focused, multi-day workshop or retreat — for instance, within the Dagstuhl or NII Shonan seminar series — would provide the ideal setting to initiate this broader discourse. Such a setting could foster deep, structured debate on the implications, methods, and governance of automation in RecSys research and lead to the co-authoring of white papers that capture and consolidate the community’s perspectives. These collective documents, in turn, could serve as catalysts for coordinated initiatives, steering both research and practice toward the systematic exploration and responsible integration of automated research systems in our field.

\section{Conclusion}
The future of recommender systems research may look very different from today’s standard practices. We stand at the cusp of a new era where AI “colleagues” could vastly accelerate the research cycle, freeing human experts to focus on high-level intuition, domain expertise, and creative vision. Embracing this future is not without challenges – there will be technical setbacks, ethical dilemmas, and perhaps scepticism from parts of our community. But the cost of inaction is high. If we ignore the trend toward research automation, we risk stagnation while other fields surge ahead. By contrast, if we actively shape and lead this trend, the RecSys community can both improve its own productivity and contribute substantially to the development of Artificial Research Intelligence.

We therefore issue this call to action as a collective invitation: Let’s create the AutoRecLab and bring automated, or even autonomous, research into the RecSys world. Let’s chart benchmarks and hold our AI agents to the same rigour we expect of human researchers. Let’s ensure transparency and ethical standards as we venture into automated science. The time is ripe – what was “wishful thinking” only yesterday is rapidly becoming an “emerging reality”\citep{beel2025evaluating}. The RecSys community must not remain on the sidelines. It is no longer sufficient to pursue incremental accuracy gains or novel algorithms alone; we should also explore innovative approaches to research. The time to engage is now!

\section*{Acknowledgements}
An initial draft of the manuscript was generated with ChatGPT using a comprehensive prompt that incorporated several research papers. The human authors performed all subsequent editing, fact-checking, and validation. We, the human authors, take full responsibility for every statement in the final text.

\bibliography{references-ais}

\begin{thebibliography}{17}
\providecommand{\natexlab}[1]{#1}
\providecommand{\url}[1]{\texttt{#1}}
\expandafter\ifx\csname urlstyle\endcsname\relax
  \providecommand{\doi}[1]{doi: #1}\else
  \providecommand{\doi}{doi: \begingroup \urlstyle{rm}\Url}\fi

\bibitem[Anand and Beel(2020)]{anand2020auto}
Rohan Anand and Joeran Beel.
\newblock Auto-surprise: An automated recommender-system (autorecsys) library with tree of parzens estimator (tpe) optimization.
\newblock In \emph{Proceedings of the 14th ACM Conference on Recommender Systems}, pages 585--587, 2020.

\bibitem[Beel et~al.(2025)Beel, Kan, and Baumgart]{beel2025evaluating}
Joeran Beel, Min-Yen Kan, and Moritz Baumgart.
\newblock Evaluating sakana's ai scientist: Bold claims, mixed results, and a promising future?
\newblock \emph{SIGIR Forum}, 59\penalty0 (1):\penalty0 1–20, October 2025.
\newblock ISSN 0163-5840.
\newblock \doi{10.1145/3769733.3769747}.
\newblock URL \url{https://doi.org/10.1145/3769733.3769747}.

\bibitem[Gottweis et~al.(2025)Gottweis, Weng, Daryin, Tu, Palepu, Sirkovic, Myaskovsky, Weissenberger, Rong, Tanno, et~al.]{gottweis2025towards}
Juraj Gottweis, Wei-Hung Weng, Alexander Daryin, Tao Tu, Anil Palepu, Petar Sirkovic, Artiom Myaskovsky, Felix Weissenberger, Keran Rong, Ryutaro Tanno, et~al.
\newblock Towards an ai co-scientist.
\newblock \emph{arXiv preprint arXiv:2502.18864}, 2025.

\bibitem[Ifargan et~al.(2024)Ifargan, Hafner, Kern, Alcalay, and Kishony]{Ifargan2024}
Tal Ifargan, Lukas Hafner, Maor Kern, Ori Alcalay, and Roy Kishony.
\newblock Autonomous {LLM}-driven research --- from data to human-verifiable research papers.
\newblock \emph{arXiv preprint arXiv:2404.17605}, 2024.

\bibitem[Jansen et~al.(2025)Jansen, Tafjord, Radensky, Siangliulue, Hope, Mishra, Majumder, Weld, and Clark]{Jansen2025}
Peter Jansen, Oyvind Tafjord, Marissa Radensky, Pao Siangliulue, Tom Hope, Bhavana~Dalvi Mishra, Bodhisattwa~P. Majumder, Daniel~S. Weld, and Peter Clark.
\newblock {CodeScientist}: End-to-end semi-automated scientific discovery with code-based experimentation.
\newblock \emph{arXiv preprint arXiv:2503.22708}, 2025.

\bibitem[Liu et~al.(2024)Liu, Liu, Zhu, Lei, Yang, Zhang, Li, and Liu]{Liu2024}
Zijun Liu, Kaiming Liu, Yiqi Zhu, Xuanyu Lei, Zonghan Yang, Zhenhe Zhang, Peng Li, and Yang Liu.
\newblock {AIGS}: Generating science from {AI}-powered automated falsification.
\newblock \emph{arXiv preprint arXiv:2411.11910}, 2024.

\bibitem[Lu et~al.(2024)Lu, Lu, Lange, Foerster, Clune, and Ha]{lu2024ai}
Chris Lu, Cong Lu, Robert~Tjarko Lange, Jakob Foerster, Jeff Clune, and David Ha.
\newblock The ai scientist: Towards fully automated open-ended scientific discovery.
\newblock \emph{arXiv preprint arXiv:2408.06292}, 2024.

\bibitem[Schmidgall and Moor(2025)]{SchmidgallMoor2025}
Samuel Schmidgall and Michael Moor.
\newblock {AgentRxiv}: Towards collaborative autonomous research.
\newblock \emph{arXiv preprint arXiv:2503.18102}, 2025.

\bibitem[Schmidgall et~al.(2025)Schmidgall, Su, Wang, Sun, Wu, Yu, Liu, Moor, Liu, and Barsoum]{Schmidgall2025}
Samuel Schmidgall, Yusheng Su, Ze~Wang, Ximeng Sun, Jialian Wu, Xiaodong Yu, Jiang Liu, Michael Moor, Zicheng Liu, and Emad Barsoum.
\newblock {Agent Laboratory}: Using {LLM} agents as research assistants.
\newblock \emph{arXiv preprint arXiv:2501.04227}, 2025.

\bibitem[Sonboli et~al.(2021)Sonboli, Mansoury, Guo, Kadekodi, Liu, Liu, Schwartz, and Burke]{sonboli2021librec}
Nasim Sonboli, Masoud Mansoury, Ziyue Guo, Shreyas Kadekodi, Weiwen Liu, Zijun Liu, Andrew Schwartz, and Robin Burke.
\newblock Librec-auto: A tool for recommender systems experimentation.
\newblock In \emph{Proceedings of the 30th ACM International Conference on Information \& Knowledge Management}, pages 4584--4593, 2021.

\bibitem[Swanson et~al.(2024)Swanson, Wu, Bulaong, Pak, and Zou]{Swanson2024}
Kyle Swanson, Wesley Wu, Nash~L. Bulaong, John~E. Pak, and James Zou.
\newblock The virtual lab: {AI} agents design new {SARS-CoV-2} nanobodies with experimental validation.
\newblock \emph{bioRxiv preprint}, 2024.

\bibitem[Tang et~al.(2025)Tang, Xia, Li, and Huang]{Tang2025}
Jiabin Tang, Lianghao Xia, Zhonghang Li, and Chao Huang.
\newblock {AI-Researcher}: Autonomous scientific innovation.
\newblock \emph{arXiv preprint arXiv:2505.18705}, 2025.

\bibitem[Vente et~al.(2023)Vente, Ekstrand, and Beel]{vente2023introducing}
Tobias Vente, Michael Ekstrand, and Joeran Beel.
\newblock Introducing lenskit-auto, an experimental automated recommender system (autorecsys) toolkit.
\newblock In \emph{Proceedings of the 17th ACM Conference on Recommender Systems}, pages 1212--1216, 2023.

\bibitem[Vente et~al.(2025)Vente, Wegmeth, and Beel]{vente2025potential}
Tobias Vente, Lukas Wegmeth, and Joeran Beel.
\newblock The potential of automl for recommender systems.
\newblock In \emph{Adjunct Proceedings of the 33rd ACM Conference on User Modeling, Adaptation and Personalization}, pages 371--378, 2025.

\bibitem[Wiggers(2025)]{heath2025aiready}
Kyle Wiggers.
\newblock Experts don't think {AI} is ready to be a co-scientist, mar 2025.
\newblock URL \url{https://techcrunch.com/2025/03/05/experts-dont-think-ai-is-ready-to-be-a-co-scientist/}.
\newblock Accessed: 2025-03-20.

\bibitem[Yamada et~al.(2025)Yamada, Lange, Lu, Hu, Lu, Foerster, Clune, and Ha]{yamada2025ai}
Yutaro Yamada, Robert~Tjarko Lange, Cong Lu, Shengran Hu, Chris Lu, Jakob Foerster, Jeff Clune, and David Ha.
\newblock The ai scientist-v2: Workshop-level automated scientific discovery via agentic tree search.
\newblock \emph{arXiv preprint arXiv:2504.08066}, 2025.

\bibitem[Young(2025)]{sakana2025aiscientistv2}
Brett Young.
\newblock Sakana unveils {AI} scientist v2, 2025.
\newblock URL \url{https://wandb.ai/byyoung3/ml-news/reports/Sakana-unveils-AI-Scientist-v2--VmlldzoxMjQxNTUzMw}.
\newblock Accessed: 2025-01-20.

\end{thebibliography}

\end{document}